\documentclass[sigconf, nonacm]{acmart}

\usepackage{algorithm}
\usepackage{algorithmicx}
\usepackage{array} % Required for the custom column type
\usepackage{booktabs}

\AtBeginDocument{%
  \providecommand\BibTeX{{%
    \normalfont B\kern-0.5em{\scshape i\kern-0.25em b}\kern-0.8em\TeX}}}

\setcopyright{acmlicensed}
\copyrightyear{2024}
\acmYear{2024}
\acmDOI{XXXXXXX.XXXXXXX}

\acmConference[Conference acronym 'XX]{Make sure to enter the correct
  conference title from your rights confirmation email}{June 03--05,
  2018}{Woodstock, NY}
\acmISBN{978-1-4503-XXXX-X/18/06}

\graphicspath{{./images/}}

\begin{document}

\title[Kernel-Based ReLU Approx. for HE-Compatible Privacy-preserving Deep Learning Models]{Kernel-Based ReLU Approximation for Homomorphic Encryption-Compatible Privacy-preserving Deep Learning Models
}

\author{Dimitrios Sygletos}
\email{d.sygletos@pasiphae.eu}
\orcid{0009-0008-5500-4048}
\affiliation{%
  \institution{Hellenic Mediterranean University}
  \city{Heraklion}
  \state{Crete}
  \country{Greece}
}

\author{Dimitra Papatsaroucha}
\email{dpapatsa@hmu.gr}
\orcid{0000-0002-8894-617X}
\affiliation{%
  \institution{Hellenic Mediterranean University}
  \city{Heraklion}
  \state{Crete}
  \country{Greece}
}

\author{Marios Choudetsanakis}
\email{m.choudetsanakis@pasiphae.eu}
\orcid{0009-0005-2210-8930}
\affiliation{%
  \institution{Hellenic Mediterranean University}
  \city{Heraklion}
  \state{Crete}
  \country{Greece}
}

\author{Ilias Politis}
\email{ilpolitis@isi.gr}
\orcid{0009-0007-2882-6452}
\affiliation{%
  \institution{Research Center "ATHENA"}
  \city{Patras}
  \state{Western Greece}
  \country{Greece}
}

\author{Evangelos K. Markakis}
\email{emarkakis@hmu.gr}
\orcid{0000-0003-0959-598X}
\affiliation{%
  \institution{Hellenic Mediterranean University}
  \city{Heraklion}
  \state{Crete}
  \country{Greece}
}

\renewcommand{\shortauthors}{Sygletos and Papatsaroucha, et al.}

\begin{abstract}
  As privacy concerns in AI technologies continue to grow, Homomorphic Encryption (HE) offers a way to perform computations on encrypted data without the need of decryption during the operations. However, HE is limited to addition and multiplication, making non-linear functions incompatible in their original form. This limitation has become more critical with the widespread use of Large Language Models (LLMs), where the non-linearity of activation functions such as the Rectified Linear Unit (ReLU) poses challenges for deployment in privacy-preserving Natural Language Processing (NLP) settings. This paper proposes a Kernel-based approximation of ReLU, enabling its use within HE-constrained settings and thus contributing a critical step toward supporting privacy-preserving LLMs. A smooth Kernel-based function, mimicking ReLU, is approximated using a second-degree polynomial, inspired by Jackson’s theorem, to achieve low multiplicative depth. The proposed method is trained and assessed directly on token embeddings from pre-trained LLMs and evaluated in various scenarios, from simulated and tokenized data to deep learning and transformer models. Results show improved approximation fidelity, supporting the method’s suitability for secure and privacy-preserving inference in various tasks.
\end{abstract}

%%
%% The code below is generated by the tool at: http://dl.acm.org/ccs.cfm



\maketitle

\section{Introduction}
In recent years, Machine Learning (ML) and Deep Learning (DL) have become integral to modern technology, powering applications from spam detection and voice recognition to recommendation systems and real-time translation. A key driver of this progress has been the ability of models to identify complex patterns in large datasets. Large Language Models (LLMs), in particular, have advanced Natural Language Processing (NLP) by using transformer architectures to generate human-like text, forming the backbone of contemporary chatbots. However, when users interact with these systems—whether for casual queries or sensitive tasks such as disclosing personal health information—their data is often transmitted and processed unencrypted. This raises privacy concerns, as model providers retain access to user inputs.

A technology that could address the concern of safeguarding data privacy during operations such as NLP is Homomorphic Encryption (HE), a Privacy-Preserving Technique (PET) enabling computations in the encrypted domain with schemes such as CKKS \cite{cheon2017ckks}. HE enables computations to be performed directly on encrypted data, without requiring decryption. However, it comes with significant limitations; most notably, that only addition and multiplication are supported, while DL models often rely on more complex, non-linear operations that are incompatible with these constraints. To convert a DL model into an HE-compatible form, activation functions like ReLU, which introduce the non-linearity essential for DL models to learn complex patterns, must be replaced with alternatives that rely solely on addition and multiplication, such as polynomials. This substitution would allow the model to operate entirely within the encrypted domain, thus preserving data privacy.

Approximating the ReLU function, however, presents a challenge due to its non-smoothness. To overcome this challenge, this paper proposes a Kernel-based smoothing technique that transforms the ReLU function into a form that can be more easily approximated by polynomials. The technique draws inspiration from Jackson’s theorem \cite{cheney1982approximation,pinkus1985nwidths}, which guarantees that smoother functions can be approximated more accurately by polynomials, with the approximation error decreasing rapidly as the polynomial degree increases. In this study, ReLU is first transformed into a smooth Kernel-based function that is then approximated using a second-degree polynomial after investigating the trade-off between approximation accuracy and computational efficiency among polynomials of various degrees. The primary objective of this work was to approximate ReLU with a HE-compatible polynomial, therefore a number of evaluation experiments were performed and the results demonstrate that the selected polynomial effectively mimics the behavior of ReLU and overcomes approximation challenges, while also maintaining compatibility with HE constraints and competitive performance in plaintext and encrypted environments among other proposed approximations identified through the literature.

The remainder of this paper is structured as follows: Section \ref{sec:Backround} provides background information about the ReLU function, Jackson's theorem, Kernel Approximations, and Polynomial Regression; Section \ref{sec:SOTA} presents relevant work that was identified across the literature regarding the approximation of ReLU using various methods; Section \ref{sec:implementation} describes the steps that were followed for implementing the proposed approximation. The setup and the evaluation of the experiments are stated in Section \ref{sec:Eval} while Section \ref{sec:Conclusion} provides the concluding remarks of this work. 

\section{Backround}
\label{sec:Backround}

\subsection{ReLU}

Activation functions are  critical components in DL models, as they introduce non-linearity, allowing neural networks to learn complex patterns  \cite{nair2010rectified}. ReLU acts as a selective gate in neural networks by preserving positive values while discarding negative ones, as can be seen in (\ref{eq:ReLU}). Such values represent messages that get transmitted from one layer of the neural network to another: when the message is deemed insignificant (negative), ReLU suppresses its transmission; otherwise, it permits it to go to the subsequent layer.

\begin{equation}
\text{ReLU}(x) = \max(0, x)
\label{eq:ReLU}
\end{equation}

This function has become the default activation function in many modern DL architectures, including neural networks and transformers, which form the foundation of modern LLMs. However, when deploying DL models in the encrypted domain, using PETs such as HE, ReLU becomes incompatible due its non-linearity. As HE supports only addition and multiplication, a practical solution is to replace ReLU with a function that can operate in the encrypted domain, such as a polynomial that includes only additions and multiplications and approximates the behavior of ReLU.

\subsection{Jackson's Theorem}

In approximation theory, Jackson's theorem relates the smoothness of a function to how well it can be approximated by polynomials, as it can be seen below. 
\begin{theorem}
    Let $f \in C^r([a, b])$, where $r \in \mathbb{N}$, and $[a, b]$ is a closed interval. Then there exists a polynomial $P_n$ of degree at most $n$ such that:
\[
\|f - P_n\|_{\infty, [a, b]} \leq \frac{C}{n^r} \|f^{(r)}\|_{\infty, [a, b]}
\]
where $\|g\|_{\infty, [a, b]} = \max_{x \in [a, b]} |g(x)|$ and $C$ is a constant depending only on $r$ and the length of the interval $(b-a)$.
\end{theorem}

This indicates that the smoother a function is—in the sense that it has more continuous derivatives—the better it can be approximated by polynomials. The error between the function and its polynomial approximation decreases rapidly as the degree $n$ of the polynomial increases. This makes Jackson’s theorem a foundational tool in approximation theory, particularly when analyzing the trade-off between approximation accuracy and polynomial complexity \cite{cheney1982approximation,pinkus1985nwidths}.

\subsection{Kernel Approximations}

Since ReLU is not smooth, directly approximating it with low-degree polynomials leads to significant error. A way to address this limitation is to first construct a smooth function that closely mimics ReLU. Kernel-based methods provide a suitable framework for generating such smooth approximations as they support learning non-linear relationships in data by implicitly mapping input features into high-dimensional spaces. Instead of explicitly transforming the data, Kernel functions compute inner products in the transformed space directly \cite{scholkopf2002learning}. 

In the context of function approximation, Kernel-based methods have demonstrated notable success in capturing non-linear patterns with a relatively small number of parameters. For example, the kernlab package \cite{kernlab} in R provides an implementation of various Kernel methods, including the hyperbolic tangent Kernel (\ref{eq:tanhdot}), which can be trained to approximate activation functions, such as ReLU. 

\subsection{Polynomial Regression}

Polynomial representations offer a way to express smooth functions under HE constraints. Polynomial regression (\ref{eq:Polynomial Regression}) offers a systematic way to approximate smooth functions with a low-degree polynomial as a form of regression analysis, in which the relationship between the independent variable $x$ and the dependent variable $y$ is modeled as a $n_{th}$  $degree$ polynomial. It extends linear regression by introducing polynomial terms to capture nonlinear patterns in the data, while still being linear in the model coefficients.

This method serves as a useful reference and approximation tool in many domains, including physics, economics, and ML, particularly when simplicity of the model and computational efficiency are important \cite{draper1998applied}. 

\section{State of the Art}
\label{sec:SOTA}

Across the literature, research has focused on approximating activation functions like ReLU using polynomials composed only of additions and multiplications. A widely used polynomial approximation of ReLU is the X-squared function showcased below in (\ref{eq:X-Squared}) \cite{tenseal,pmlr-v48-gilad-bachrach16}.

\begin{equation}  
    f(x) = x^2 \label{eq:X-Squared}
\end{equation}

\begin{figure}[H]
    \centering
    \includegraphics[width=0.8\linewidth]{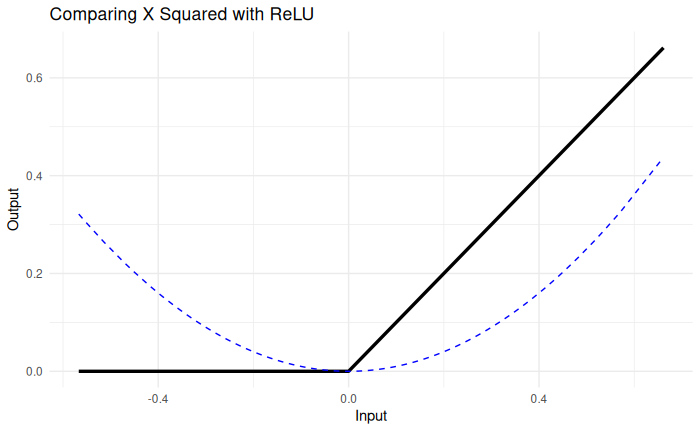}
    \caption{Actual ReLU vs X Squared}
    \label{fig:ReLu Approx.}
    \Description{ReLu Approx.}
\end{figure}

While this function is compatible with HE, it lacks the zeroing behavior of ReLU for negative inputs. As can be seen in Figure \ref{fig:ReLu Approx.}, the X-squared function maps all inputs, both positive and negative, to positive values, thereby altering the nature of the activation and introducing unintended signal propagation to the subsequent layers.

HE libraries like TenSEAL \cite{tenseal} embed the X-squared function within CNNs and \cite{lecun1998gradient} evaluated its efficacy based on the prediction results on synthetic datasets, such as MNIST \cite{mnist}. However, this methodology assessed the approximation indirectly, as it compared classification accuracy of the CNN model, which may not reflect the quality of the approximation itself.

In \cite{juvekar2018faster} low degree polynomials were proposed, capturing the behavior to common activation functions (e.g., Swish, Softplus, and ReLU). The proposed Minimax method (\ref{eq:FasterCryptoNets Polynomial}) performed well in the image classification task using the MNIST dataset, showing that it could replace the original activation functions; however, with a significant disadvantage as its coefficients are arbitrary real numbers that do not allow for efficient encrypted computation.
\begin{equation}
    \setlength{\abovedisplayskip}{1pt} % Reduces space above
    \setlength{\belowdisplayskip}{1pt} % Reduces space below
    p(x) = 0.125x^2 + 0.5x + 0.25
    \label{eq:FasterCryptoNets Polynomial}
\end{equation}

An alternative approach utilized Chebyshev polynomials in \cite{10538562}. While this method provided good approximations, it required polynomials of degree at least three. According to Jackson's theorem, directly approximating non-smooth functions, such as ReLU, requires higher degree polynomials to achieve acceptable accuracy \cite{cheney1982approximation}. Thereby, the reviewed study used polynomials of the 7th and 9th degree, which on one hand optimized the approximation but on the other would increase the computational cost in an HE scheme. 

As revealed through the literature, most studies have focused on approximating ReLU through functions that either fall short in preserving its essential behavior—as the goal is to approximate a function that is not $ C^\infty$ (not differentiable at zero)—or evaluated the performance of their proposed approaches within specific model architectures rather than assessing their fidelity on real-world data. The approach proposed in this paper puts forward a Kernel-based method that forms the basis for producing a polynomial approximation of ReLU, designed considering HE constraints and computation costs, which was evaluated by training on real embedding data from pre-trained LLMs and tested in different scenarios.

\section{Implementation} \label{sec:implementation}
In this research, a second-degree polynomial is proposed to be fitted in order to minimize the combinatorial explosion and prohibitive expenses related to high-degree polynomial Kernel mappings. High dimensionality significantly constrains the practical deployment of high-degree Polynomial Kernels, despite their expressive power. The feature space grows at a rate of $O({n^d})$ as the polynomial degree \textit{d} increases, leading to a substantial increase in the number of terms. According to Ahle et al. \cite{ahle2020oblivioussketchinghighdegreepolynomial}, the resulting Kernel matrices become so dense that downstream primitives, such as Kernel PCA or Ridge Regression, require quadratic space and time complexity, which can be prohibitive for large-scale datasets, leading standard Kernel methods to lose their computational efficiency.

The sub-sections that follow delve into the details of the implementation process regarding input data selection, kernel regression, input data selection, and polynomial regression. In addition, polynomials of various degrees were investigated prior to the second-degree polynomial selection. The results of this investigation are also provided in the following sections, including the final polynomial that was obtained through the process for effectively mimicking ReLU considering the core objective of this study to support privacy-preserving LLMs.

\subsection{Kernel Regression}

As already mentioned, the main difference from applying a polynomial regression directly to ReLU is the smoothness of the function. Since ReLU is not $ C^\infty$, a more accurate approximation can be obtained by first replacing it with a smooth function. The kernlab package \cite{kernlab} provides various kernel-based ML methods for approximations of functions, assisting in the design of a kernel function mimicking the ReLU function in this work. The kernel function that was trained on the input data was the hyperbolic tangent kernel tanhdot (\ref{eq:tanhdot}).

\begin{equation}
    k(x, x') = \tanh( \langle x, x' \rangle + 1)
    \label{eq:tanhdot}
\end{equation}

\begin{figure}[H]
    \centering
    \includegraphics[width=0.8\linewidth]{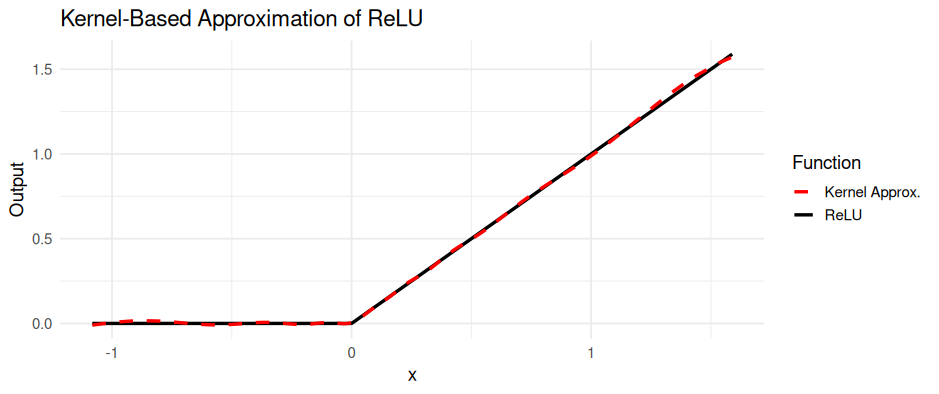}
    \caption{Actual ReLU vs Kernel Approx.}
    \label{fig:Kernel Validation}
    \Description{Kernel Validation}
\end{figure}

As can be seen in Figure \ref{fig:Kernel Validation}, this kernel function was selected because the approximation of ReLU is significant, capturing entirely both positive and negative values, where the negative ones are replaced with zero \cite{nair2010rectified}.

\subsection{Input Data Selection, Polynomial Regression \& Optimal Degree Selection}

To approximate the output of the kernel function, a polynomial regression (\ref{eq:Polynomial Regression}) model was trained on the kernel function predictions, where  $ n $ is the degree of the polynomial. 

\begin{equation}
    y = \beta_0 + \beta_1x + \beta_2x^2 + ... + \beta_nx^n + \epsilon
    \label{eq:Polynomial Regression}
\end{equation}

Regarding input data, embeddings were extracted from RoBERTa\footnote{https://huggingface.co/FacebookAI/roberta-base} and DistilBERT\footnote{https://huggingface.co/distilbert/distilbert-base-uncased} using the Stanford Sentiment Treebank (SST-2) dataset \cite{socher2013recursive}, a widely used benchmark for sentence-level sentiment classification, where each sentence is labeled as either positive or negative. The two LLM models were utilized in their pre-trained state without architectural changes, configured with a batch size of 32 and a max length of 128 tokens. These hyperparameter values reflect a typical industry trade-off that maintains a balance between the requirement for reliable gradient updates and adequate sequence context for successful model convergence, as well as hardware memory efficiency and computational performance. Furthermore, SST-2 was selected due to its standardized structure and demonstrated dependability in capturing an array of linguistic attributes. SST-2 is built from movie reviews but its sentences are short, varied, and linguistically rich, which makes it a standard choice for evaluating language models and transformer architectures. Because SST-2 is part of the GLUE benchmark suite, strong performance on it is often reported in studies as evidence of a model’s general NLP capability, such as \cite{wang2019glue, devlin2019bert}. Moreover, the inclusion of both RoBERTa and DistilBERT serves a dual purpose: the former represents the upper bound of Transformer-based accuracy through optimized pre-training, while the latter provides insight into model generalization under architectural constraints. This selection ensures that the findings of this study are representative of both high-capacity and resource-optimized models.

A non-linear relationship was fitted between the predictions of the kernel function and the actual tokenized vector of the validation set of SST-2, through polynomial regression. The coefficients were estimated using the least-squares method. Regarding optimal polynomial degree selection, polynomials of degrees 2, 3, 4, and 5 were defined and subsequently integrated as a ReLU replacements in plaintext and encrypted environments. These were developed as surrogate models, fitted via Ordinary Least Squares (OLS) to the predictive outputs of a Tanh-kernel Gaussian Process to ensure accuracy during evaluation. Testing was performed on a system equipped with an NVIDIA H100 GPU\footnote{https://www.nvidia.com/en-eu/data-center/h100/}, which was used to accelerate the kernel fitting and polynomial regression steps, while all HE-related computations and performance measurements were executed on the host CPU. Furthermore, encrypted evaluations were performed under the CKKS HE scheme. The polynomials of various degrees were evaluated based on two competing objectives:

\begin{itemize}
\item Approximation Accuracy: Minimizing the error between the polynomial and the ReLU function (i.e., how closely the polynomial mimics the original ReLU function) to preserve the model’s decision-making capabilities.

\item Computational Complexity: Minimizing the computational weight of the operation, specifically multiplicative depth and latency, to ensure the model remains efficient.
\end{itemize}

With regard to the first objective, the evaluation was centered around the Mean Square Error (MSE) (\ref{eq:MSE}) to assess approximation accuracy. MSE quantifies the average squared difference between predicted and true values, where a lower MSE value indicates better polynomial performance, as it reflects smaller deviations between predictions and true values.

\begin{equation}
\text{MSE} = \frac{1}{n} \sum_{i=1}^{n} (y_i - \hat{y}_i)^2
\label{eq:MSE}
\end{equation}

Regarding the second objective, the performance was further evaluated based on computational performance and resource usage to identify the viability of the operation. This involved the measurement of inference time (i.e., time required to apply the polynomial to perform the prediction), overall training time, and memory usage on unencrypted and encrypted data. These metrics were used to provide a holistic picture of the trade-offs between approximation accuracy and the hardware requirements needed for model deployment.

Contrary to the expectation that higher-degree polynomials would provide a better fit, it is observed in Table \ref{Approximation Accuracy by Polynomial Degree} that lower-degree polynomials provide a significantly more stable approximation. Degree-2 polynomial achieved the lowest MSE in both plaintext and encrypted environments, with a value of 0.056862 in both cases, while higher-order polynomials, particularly Degree-4 and Degree-5, exhibited significant instability when approximating the ReLU function within the target range, leading to exponentially higher error rates. Notably, the transition to the encrypted domain introduced negligible error (less than 10$^{-5}$ deviation), confirming the stability of the proposed polynomial implementation within the CKKS scheme.

\begin{table}[h!]
\centering
\caption{Approximation Accuracy by Polynomial Degree}
\label{Approximation Accuracy by Polynomial Degree}
\begin{tabular}{lcc}
\toprule
\textbf{Polynomial Degree} & \textbf{Plaintext MSE} & \textbf{Encrypted MSE}\\
\midrule
Degree-2   & 0.056862 & 0.056862 \\   
Degree-3   & 0.784026 & 0.784037 \\ 
Degree-4   & 4.909682 & 4.909823 \\
Degree-5   & 390.4527 & 390.4659 \\

\bottomrule
\end{tabular}
\end{table}

Furthermore, the computational performance and memory usage footprints of polynomials across different degrees are detailed in Table \ref{Computational Performance and Resource Footprint}, where it is observed that Encrypted Inference Time is substantially increased as the polynomial degree rises. Notably, the Degree-5 inference duration (0.068582 s) is found to be approximately 14.7 times slower than the Degree-2 inference (0.004646 s), representing a critical bottleneck in HE, as increased multiplicative depth accelerates noise growth. Similarly, the Total Time metric, encompassing the entire pipeline including the overhead of cryptographic setup, encryption, and decryption, showcases an increase as the polynomial degree rises. However, a surprising decline in memory usage is demonstrated for the 4th and 5th polynomial degrees. It would be expected that resource consumption would also increase as the polynomial degree rises, which is the case when comparing the 2nd and 3rd polynomial degrees in terms of memory usage. This counterintuitive trend is likely attributable to implementation-level factors, such as memory allocation and reuse patterns within the runtime environment, rather than a systematic reduction in memory requirements as polynomial degree increases. Although a reduction of 0.002 MB in memory usage is logged for Degree-5, Degree-2 is identified as the actual winner for efficiency. Superior accuracy is achieved by Degree 2 in 1/9th of the time, which would allow for much higher throughput and lower total hardware costs in a production environment.

\begin{table}[h!]
\centering
\caption{Computational Performance and Resource Footprint}
\label{Computational Performance and Resource Footprint}
\setlength{\tabcolsep}{1pt} % Adjusts spacing between columns
\begin{tabular}{lccccc}
\toprule
\textbf{Degree} & \textbf{Encr. Inf. Time (s)} & \textbf{Total Time (s)} & \textbf{ Memory (MB)} \\
\midrule
Degree-2   & 0.004646     & 0.172997       & 0.154295      \\   
Degree-3   & 0.022623     & 0.583531       & 0.154436      \\ 
Degree-4   & 0.046998     & 1.168747       & 0.152130      \\
Degree-5   & 0.068582     & 1.515629       & 0.151991     \\
\bottomrule
\end{tabular}
\end{table}

Based on the evidence provided by the results of the investigation, the Degree-2 polynomial was selected as the final kernel, as it provides the optimal trade-off by delivering the most accurate approximation of the ReLU function among the various-degree polynomials that were tested, while maintaining the lowest computational footprint, establishing an acceptable balance among expressive capability and computational limits. This choice ensures that the solution remains efficient even when scaled to complex neural network architectures in the encrypted domain.

\subsection{Algorithm \& Final Output}

The following algorithm details the process described in the above sub-sections, for approximating the ReLU function by first training a Kernel function on embedding data, and then fitting a second-degree polynomial to its outputs. 

\begin{algorithm}[H]
\caption{ReLU Approximation via Kernel and Polynomial Fitting}
\label{alg:relu_kernel_poly}
\begin{algorithmic}[1]
\State Let $\mathcal{D}_{\text{train}} = \{(x_i, y_i)\}_{i=1}^N$, 
$\mathcal{D}_{\text{val}} = \{(x_j, y_j)\}_{j=1}^M$, and 
$\mathcal{D}_{\text{test}} = \{(x_k, y_k)\}_{k=1}^T$, where $y = \text{ReLU}(x) = \max(0, x)$.
\Statex

\State \textbf{Kernel Training:} Train a Kernel function $K(x, x')$ to approximate the ReLU function by minimizing:
\[
\min_{\alpha} \; \mathcal{L}_{\text{Kernel}} = \frac{1}{N} \sum_{i=1}^{N} \big( K(x_i; \alpha) - y_i \big)^2,
\]
where $\alpha$ are the Kernel parameters.
\Statex

\State \textbf{Kernel Prediction:} Use the trained kernel to predict outputs on the validation set:
\[
\hat{y}_j = K(x_j; \alpha^*), \quad \forall (x_j, y_j) \in \mathcal{D}_{\text{val}}
\]
\Statex

\State \textbf{Polynomial Fitting:} Fit a second-degree polynomial 
$p(x) = a_0 + a_1x + a_2x^2$ 
to the Kernel predictions by minimizing:
\[
\min_{a_0, a_1, a_2} \; \mathcal{L}_{\text{poly}} = \frac{1}{M} \sum_{j=1}^{M} \big( p(x_j) - \hat{y}_j \big)^2
\]
\Statex

\State \textbf{Final Approximation:} 
The approximated ReLU function is given by:
\[
\widetilde{\text{ReLU}}(x) = a_0 + a_1x + a_2x^2
\]
\Statex
\end{algorithmic}
\end{algorithm}

After applying all the steps of the methodology, the resulting polynomial obtained—hereinafter Kernel Polynomial method—is the one presented in (\ref{eq:Kernel Polynomial}) below:

\begin{equation}
    p(x) = 0.082261 + 0.495588x + 0.444488x^2
    \label{eq:Kernel Polynomial}
\end{equation}

\section{Evaluation}
\label{sec:Eval}
Given that the primary objective of this work was to approximate ReLU with a HE-compatible polynomial, the evaluation encompasses a comprehensive analysis, including MSE assessments as well as  model performance and computational cost measurements. The proposed Kernel Polynomial method is compared against existing benchmarks from the literature (described in Section \ref{sec:SOTA}) with regard to both approximation accuracy and performance when applied to real-world challenges in both plaintext and encrypted environments.

\subsection{Experimental Design}
The evaluation consists of six experimental scenarios, each designed to isolate and evaluate specific aspects of the proposed Kernel Polynomial method. A shared testbed and consistent evaluation protocol are maintained across all scenarios to ensure comparability.

\begin{itemize}
    \item Experiments 1 \& 2 – Polynomial Approximation Performance: Using tokenized data as a consistent framework, polynomial approximations are evaluated on tokenized data to establish a baseline under controlled conditions (Experiment 1), and subsequently applied to structured data to evaluate the robustness and generalization when handling complex, structured inputs (Experiment 2).
    \item Experiments 3 \& 4 - Model Performance: These experiments focus on downstream impact. In Experiment 3, the MNIST dataset is used to measure the predictive accuracy of two DL models, determining how approximations affect standard neural network architectures. In Experiment 4, the assessment is extended to a transformer-based model, evaluating its performance and adaptability across two district datasets: CIFAR-10 and CIFAR-100.
    \item Experiment 5 - Computational Efficiency: The computational cost of the approximations is evaluated across all scenarios. Factors such as runtime and resource utilization are isolated while model and data conditions remain fixed.
    \item Experiment 6 - Encrypted Environment Performance: This experiment evaluates the approximations' transition to an encrypted environment, quantifying the trade-off between classification accuracy and the computational slowdown caused by cryptographic overhead to validate the practical viability of the approximations.

\end{itemize}

For polynomial approximation performance, evaluation was centered around MSE as defined earlier in (\ref{eq:MSE}) while for the DL and transformer models the accuracy metric was used, as defined in (\ref{eq:acc}), as a straightforward measure of overall classification performance by comparing predicted labels with ground-truth targets.

\begin{equation}
    \text{Accuracy} = \frac{\text{Number of Correct Predictions}}{\text{Total Number of Predictions}}
    \label{eq:acc}
\end{equation}

\subsection{Experiment Environment}

The transformers library \cite{transformers} was used to tokenize and convert textual input into vector representations. The torch package in Python\footnote{https://www.python.org/} was used to implement and train the DL models, including the Vision Transformer (ViT). Additionally, the NVIDIA H100 GPU was utilized also for the evaluation process of this study, solely to accelerate polynomial regression and kernel fitting in order to establish a consistent benchmarking environment. On the other hand, all HE-related computations were strictly restricted to the host CPU since standard schemes like CKKS currently lack native GPU acceleration. This approach was utilized to ensure a fair comparison between encrypted and plaintext processing, whereby the precise computational overhead is presented without hardware-induced bias. By focusing on memory usage metrics, a transparent analysis of the cryptographic overhead and resource consumption was achieved without the confounding factors of non-standardized GPU acceleration. Consequently, the performance results characterize the system's memory usage, providing an accurate reflection of high computational demands. 

\subsection{Evaluation Setup}

\subsubsection{Independent Variables}
The features that were consciously adjusted in each of the six experimental scenarios indicate the independent variables in this study. These include the approach to modeling used (polynomial approximation methods, DL models, and a transformer-based model), the type of data used for evaluation (tokenized and structured data), and the specific evaluation focus of each experiment, such as approximation accuracy, model prediction performance, and computational efficiency. A controlled comparison between various modeling and data conditions is possible since each experimental scenario ensures a consistent testbed while representing a unique configuration of these parameters.

\subsubsection{Dependent Variables}
The defined independent variables have a direct impact on the dependent variables, and they correspond to the performance results measured under each experimental scenario. These refer to quantitative performance metrics such as MSE for evaluating the efficiency of the polynomial approximations, accuracy metrics for evaluating the predictive performance of the DL and transformer models, and computational cost measurements, including runtime and resource utilization. These metrics allow an unbiased assessment of each modeling approach's performance under the specified experimental conditions.

\subsection{Results}
\subsubsection{Experiment 1: Baseline Approximation Performance}

A series of tests was conducted using multiple sample sets drawn from normal distributions to evaluate the stability and robustness of the various approximations. By generating independently sampled inputs with varying random seeds, the aim was to assess whether performance remained consistent across different distributions.

The results in Table \ref{tab1:norm results} depict the MSE for the various ReLU approximation methods, included in each row, tested against data samples from Normal Distributions with increasing levels of variance, showcased in each column. The data revealed that the proposed Kernel Polynomial method is the most accurate in low-to-moderate noise scenarios, while FasterCryptoNets demonstrates greater resilience only under the highest variance perturbations (N(0,1)). Nevertheless, the Kernel Polynomial remained the most stable and reliable approximation overall, as its variance consistently remained below 0.5.

\begin{table}[h!]
\centering
\caption{MSE of different ReLU approximation methods on vectors from various samples of Normal distribution.}
\label{tab1:norm results}
\setlength{\tabcolsep}{2.5pt} % Adjusts spacing between columns
\begin{tabular}{lccccc}
\toprule
\textbf{Approximation} & \textbf{N(0,0.1)} & \textbf{N(0,0.2)} & \textbf{N(0,0.5)} & \textbf{N(0,0.7)} &\textbf{N(0,1)} \\
\midrule
Kernel Polynomial   & \textbf{0.002}     & \textbf{0.001}     & \textbf{0.003}   & \textbf{0.018}   & 0.147  \\   
X Squared           & 0.003     & 0.012       & 0.113   & 0.418   & 1.904  \\ 
FasterCryptoNets    & 0.045     & 0.033       & 0.018   & \textbf{0.018}   & \textbf{0.022}  \\
Chebyshev - 3       & 1.726     & 1.628       & 1.371   & 1.228   & 1.048  \\
Chebyshev - 5       & 0.679     & 0.620       & 0.478   & 0.407   & 0.327  \\
\bottomrule
\end{tabular}
\end{table}

\subsubsection{Experiment 2: Approximation Performance on Structured Data}

To ensure a realistic evaluation, the SST-2 dataset was derived from different pre-trained LLMs (RoBERTa and DistilBERT). The input to each polynomial consisted of the test set including tokenized text values that were not used in the training process. The output corresponded to either the true values of the ReLU activation function or the predictions generated by different approximations. As it can be seen in Table \ref{tab2:tokenized results}, it was concluded that the proposed Kernel Polynomial method outperformed the other approximations, achieving the lowest MSEs in both cases. It can also be observed that Kernel Polynomial showcases a better performance compared to the FasterCryptoNets method that performed slightly better in the previous experiment, while Chebyshev-3 and -5 methods provided the highest MSEs across all approximation methods.

\begin{table}[h!]
\centering
\caption{MSE of different ReLU approximation methods on tokenized vectors from RoBERTa and DistilBERT.}
\label{tab2:tokenized results}
\begin{tabular}{lcc}
\toprule
\textbf{Approximation} & \textbf{RoBERTa-MSE} & \textbf{DistilBERT-MSE} \\
\midrule
Kernel Polynomial   & \textbf{0.002}     & \textbf{0.002}\\   
X Squared           & 0.016     & 0.033\\ 
FasterCryptoNets    & 0.037     & 0.024\\
Chebyshev - 3       & 1.641     & 1.500\\
Chebyshev - 5       & 0.629     & 0.548\\
\bottomrule
\end{tabular}
\end{table}

\begin{figure}[H]
    \centering
    \includegraphics[width=0.9\linewidth]{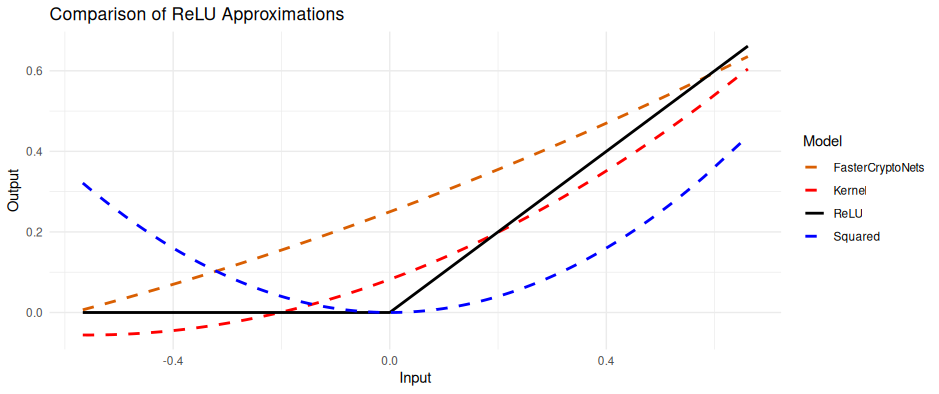}
    \caption{Actual ReLU vs Approximations.}
    \label{fig:Evaluation of Approximations}
    \Description{Evaluation of Approximations}
\end{figure}

Moreover, Figure \ref{fig:Evaluation of Approximations} presents a graphical representation of the approximations, including the proposed Kernel Polynomial method. On the x-axis, the tokenized vectors are displayed, while on the y-axis, the corresponding output values of each approximation are shown. The solid black line represents the ReLU function. As can be observed, the X-squared function inadequately accounts for negative values and is very far from positive values. Furthermore, the FasterCryptoNet polynomial exhibits linear behavior and does not account for negative values. The red dashed line represents the proposed Kernel Polynomial method that more accurately approximates ReLU. The Chebyshev methods, which exhibited significantly higher MSE, were excluded from the figure for clarity.

\subsubsection{Experiment 3: Deep Learning Models Performance}

The proposed Kernel Polynomial method was assessed alongside the established polynomials from the literature, which were also included in the previous experiments, to determine if model accuracy is maintained when the proposed solution is implemented.

A variety of neural network structures constitute the fundamental basis of DL models, providing the foundational mechanisms for learning hierarchical and nonlinear representations from data, enabling a wide range of supervised learning tasks. In this work, FeedForward and Convolutional neural networks (CNNs) were selected for conducting Experiment 3. Feedforward neural networks were used as a baseline due to their simplicity and suitability for supervised classification and regression tasks, while CNNs were selected for their effectiveness in capturing hierarchical spatial features through weight sharing and local receptive fields.

The results presented in Table \ref{tab5:deep learning results} demonstrate that the proposed Kernel Polynomial method achieved a high degree of prediction accuracy, comparable to established polynomial methods. Particularly, in the FeedForward model, an accuracy of 99\% is achieved by the Kernel Polynomial, which is identical to the performance of X Squared, while both were slightly outperformed by FasterCryptoNets, by a narrow margin of 0.3\%. In addition, the Kernel Polynomial method demonstrated the highest performance (99.2\%) when employed in the CNN model. The lowest accuracy levels across both architectures were consistently produced by Chebyshev-based approximations. It is confirmed by these findings that a highly effective formulation is provided by the Kernel Polynomial method despite its simpler form, maintaining competitive performance against more complex state-of-the-art approximations.

\begin{table}[h!]
\centering
\caption{Accuracy of the DL models}
\label{tab5:deep learning results}
\begin{tabular}{lcc}
\toprule
\textbf{Approximation} & \textbf{FeedForward} & \textbf{CNN} \\
\midrule
Kernel Polynomial   & 99\%     & \textbf{99.2\%} \\   
X Squared           & 99\%     & 99.1\% \\ 
FasterCryptoNets    & \textbf{99.3\%}     & 99.1\% \\
Chebyshev - 3       & 98.7\%     & 98.4\% \\
Chebyshev - 5       & 98.9\%     & 97.8\% \\
\bottomrule
\end{tabular}
\end{table}

\subsubsection{Experiment 4: Transformer Model Performance}

The Vision Transformer (ViT) adapts the transformer architecture from NLP to computer vision by treating images as sequences of patches rather than spatial grids. Instead of relying on convolutional operations, ViT learns global relationships through self-attention, enabling it to model long-range dependencies across an image.

To evaluate and utilize the capabilities of the transformer architecture, the CIFAR-10 \cite{Krizhevsky2009LearningML} and CIFAR-100 \cite{Krizhevsky2009LearningML} datasets were utilized. These datasets, developed at the University of Toronto\footnote{https://www.utoronto.ca/} as part of the Canadian Institute for Advanced Research (CIFAR\footnote{https://cifar.ca/}), are considered recognized benchmarks to evaluate a transformer's ability to generalize across both general and detailed image classification tasks. The various levels of complexity provide an effective basis for evaluating the effectiveness of the model while ensuring that the results remain directly comparable with well-established literature in the field.

As shown in Table \ref{tab6:transformer results}, the suggested Kernel Polynomial method achieved performance comparable to other polynomial approximations within the ViT transformer architecture. The X Squared approximation had a slightly higher precision of 65.1\% on CIFAR-10, marginally outperforming the Kernel Polynomial by 0.4\%. On the other hand, the Kernel Polynomial showed better generalization on the CIFAR-100 dataset, with the highest accuracy of 38.5\%. In addition, it is worth noting that higher-degree approximations, such as Chebyshev and FasterCryptoNets, did not demonstrate greater accuracy.

\begin{table}[h!]
\centering
\caption{Accuracy of the ViT transformer}
\label{tab6:transformer results}
\begin{tabular}{lcc}
\toprule
\textbf{Approximation} & \textbf{CIFAR 10} & \textbf{CIFAR 100} \\
\midrule
Kernel Polynomial   & 64.7\%     & \textbf{38.5\%} \\   
X Squared           & \textbf{65.1\%}     & 38.1\% \\ 
FasterCryptoNets    & 64.2\%     & 37.7\% \\
Chebyshev - 3       & 60.3\%     & 33.2\% \\
Chebyshev - 5       & 62.0\%     & 35.6\% \\
\bottomrule
\end{tabular}
\end{table}

\subsubsection{Experiment 5: Computational Efficiency}

The computational cost is the quantity of resources required for the completion of a computer operation. Variables such as the training time needed to train and evaluate a model are often interpreted as computational costs. Computational cost is crucial as it has an immediate effect on an approach’s feasibility, efficiency, and scalability, especially when it involves large datasets or complex algorithms. In this experiment, the training cost, measured in execution time (in seconds), was evaluated on the CIFAR-10 and CIFAR-100 datasets to assess the computational overhead associated with each approximation method. 

The computational cost of training DL models using diverse approximations are demonstrated in Table \ref{tab8:Computational DL}, showcasing that the X Squared approximation achieved the lowest total latency, encompassing both the training phase and the inference time on the test partition. This can be explained by its simple arithmetic polynomial form that only requires element-wise addition and multiplication, avoiding more computationally costly nonlinear operations, such as divisions or exponentials. Notably, even though the proposed Kernel Polynomial was outperformed by X Sqaured in this experiment, a significant difference in training time is observed when the Kernel Polynomial approximation is compared to higher-degree polynomials, highlighting the efficiency of the proposed method despite being only a second-degree polynomial.

\begin{table}[h!]
\centering
\caption{Computational Cost of the DL models }
\label{tab8:Computational DL}
\begin{tabular}{lcc}
\toprule
\textbf{Approximation} & \textbf{FeedForward} & \textbf{CNN} \\
\midrule
Kernel Polynomial   & 100 \text{sec} & 399 \text{sec} \\  
X Squared           & \textbf{91 \text{sec}} & \textbf{357 \text{sec}}\\ 
FasterCryptoNets    & 103 \text{sec} & 409 \text{sec} \\
Chebyshev - 3       & 107 \text{sec} & 560 \text{sec} \\
Chebyshev - 5       & 118 \text{sec} & 956 \text{sec} \\
\bottomrule
\end{tabular}
\end{table}

Furthermore, the results in Table \ref{tab7:Computational ViT} indicate that the Kernel Polynomial method required the lowest training time for ViT transformers on both CIFAR-10 and CIFAR-100, with  936 s and 931 s, respectively. Comparable performance was observed for the X Squared approximation, with training times of 937 s on CIFAR-10 and 935 s on CIFAR-100. These two approaches demonstrated a clear computational advantage over the remaining approximation methods. In particular, FasterCryptoNets exhibited higher training times on CIFAR-10 and CIFAR-100, with 944 s and 945 s, respeticely, while Chebyshev-3 required 960 s and 950 s. The highest training cost was observed for Chebyshev-5, with execution times of 1094 s on CIFAR-10 and 1091 s on CIFAR-100. 

\begin{table}[h!]
\centering
\caption{Computational Cost of the ViT transformers }
\label{tab7:Computational ViT}
\begin{tabular}{lcc}
\toprule
\textbf{Approximation} & \textbf{CIFAR 10} & \textbf{CIFAR 100} \\
\midrule
Kernel Polynomial   & \textbf{936 \text{sec}} & \textbf{931 \text{sec}} \\  
X Squared           & 937 \text{sec} & 935 \text{sec}\\ 
FasterCryptoNets    & 944 \text{sec} & 945 \text{sec} \\
Chebyshev - 3       & 960 \text{sec} & 950 \text{sec} \\
Chebyshev - 5       & 1094 \text{sec} & 1091 \text{sec} \\
\bottomrule
\end{tabular}
\end{table}

\subsubsection{Experiment 6: Encrypted Environment Performance}

This experiment was conducted to serve as the ultimate validation of the proposed Kernel Polynomial method by transitioning from a plaintext baseline to a fully homomorphic encryption environment using the CKKS scheme and the ViT transformer. The experiment focused on two primary objectives: i) the comparison of the classification accuracy of the ViT model before and after encryption, with the goal to observe the "Accuracy Gap" caused by the conversion of weights and activations into the encrypted domain, depicted in Table \ref{Classification Accuracy Gap (Plaintext vs. Encrypted MNIST)}, and ii) the comparison of the Slowdown Factor, defined as the computational expansion and latency increase caused by cryptographic overhead through the measurement of the total time needed to complete the tasks in the encrypted domain compared to the plaintext baseline, depicted in Table \ref{Computational Efficiency and Latency Expansion}, to highlight the cost of privacy. These metrics were benchmarked exclusively on the host CPU, despite the use of the NVIDIA H100 system employed for all experiments, to account for the fact that HE-based computations are strictly CPU-bound. Through this experiment, the practical trade offs required to move from theoretical approximation to a deployable, privacy preserving DL solution are isolated, while assessing how the mathematical stability of the second-degree kernel-based polynomial impacts classification accuracy in a high privacy setting.

\begin{table}[h!]
\centering
\caption{Classification Accuracy Gap (Plaintext vs. Encrypted MNIST)}
\label{Classification Accuracy Gap (Plaintext vs. Encrypted MNIST)}
\setlength{\tabcolsep}{2pt} % Adjusts spacing between columns
\begin{tabular}{lccccc}
\toprule
\textbf{Model Approach} & \textbf{Plaintext Acc} & \textbf{Encrypted Acc} & \textbf{Acc. Gap (\%)} \\
\midrule
Kernel Polynomial   & \(97.3\%\)    & \textbf{68.0\%\ }     & \textbf{-29.3\%\ }  \\   
X Squared           & \(97.4\%\)    & \textbf{68.0\%\ }    & \(-29.4\%\)   \\ 
FasterCryptoNets    & \(97.3\%\)    & \(66.0\%\)    & \(-31.3\%\)   \\
Chebyshev - 3       & \(97.3\%\)    & \(62.0\%\)     & \(-35.3\%\)  \\
Chebyshev - 5       & \textbf{97.6\%\ }    & \(64.0\%\)     & \(-33.6\%\)  \\
\bottomrule
\end{tabular}
\end{table}

Despite having similar performance in plaintext, the accuracy of higher-degree approximations, like Chebyshev-3, dropped significantly in the encrypted domain compared to the proposed Kernel Polynomial, with 62.0\% and 68.0\%, respectively. In addition, the Kernel Polynomial exhibited the lowest Accuracy Gap compared to all approximations, at 29.3\%, suggesting that the added complexity of higher degree polynomial introduces more noise than benefit within the CKKS context.

\begin{table}[h!]
\centering
\small
\caption{Computational Efficiency and Latency Expansion}
\label{Computational Efficiency and Latency Expansion}
\setlength{\tabcolsep}{2pt} % Keep columns tight to fit the page
\begin{tabular}{l cccc}
\toprule
\textbf{Model Approach} & 
\textbf{\shortstack{Total Exp. \\ Time (s)}} & 
\textbf{\shortstack{Enc. Train \\ Time (s)}} &
\textbf{\shortstack{Slowdown \\ Factor}} &
\textbf{\shortstack{Peak Memory \\ Usage (MB)}} \\
\midrule
Kernel Polynomial   & \textbf{1864.19}  & \textbf{1597.46}  & \textbf{18.5$\times$}   & \textbf{7.83}\\    
X Squared            & 1872.86  & 1605.48  & 18.6$\times$  & 7.94\\ 
FasterCryptoNets     & 1917.99  & 1644.31  & 19.0$\times$  & 7.81\\
Chebyshev - 3        & 6780.00  & 5808.09  & 67.3$\times$  & 7.81\\
Chebyshev - 5        & 11348.86 & 9730.65  & 109.7$\times$ & 7.81 \\
\bottomrule
\end{tabular}
\end{table}

Regarding the cost of privacy, the experiment results showcase that replacing a second-degree polynomial with a higher order polynomial significantly raises latency. The models integrating the Chebyshev-3 and Chebyshev-5 approximations were extremely slow compared to model employing the proposed Kernel Polynomial, requiring 6780s and 11348s, respectively, while the proposed method achieved 1864 s of total experiment time, encompassing the entire pipeline of training, inference, and the models encryption-decryption processes. This demonstrated that as the polynomial degree rises, it becomes significantly more challenging to operate with in an encrypted setting. Notably, the model integrating the Chebyshev-5 polynomial demonstrated an increase in total experiment duration by over 109 times. Additionally, roughly the same amount of memory (7.8 MB) was consumed by all models across all approximations, while only the model employing the X Square approximation required more than 7.9 MB. It is worth noting that the significant disparities in execution time confirm that latency, rather than memory, constitutes the primary bottleneck. Consequently, the suggested Kernel Polynomial can be considered the most effective approach to high-throughput, privacy-preserving applications considering the exponential costs of the higher-degree approximations.

During this experiment, the Kernel Polynomial performed more effectively than higher-degree approximations since it maintained an appropriate balance between mathematical stability and the hardware limits of the encrypted domain. Higher-order polynomials, such as the Chebyshev-5, should fit more effectively in theory, however they are prone to Runge's phenomenon \cite{FORNBERG2007379}, according to which the use of a high-degree polynomial to interpolate a function over equally spaced points causes the error to increase significantly, resulting in extreme oscillations near the edges of the interval. This can also be observed in Table \ref{Approximation Accuracy by Polynomial Degree} provided earlier in Section \ref{sec:implementation}, where it is depicted that the MSE rises from 0.056 to over 390 as the polynomial degree increases in encrypted environments. Additionally, the fact that higher-degree polynomial operations require a higher multiplicative depth accelerates the noise development in the CKKS encryption scheme, which lowers the signal-to-noise ratio and makes it harder to classify correctly. The linear derivative of the proposed 2nd-degree Kernel Polynomial ensures stable gradient flow from a training perspective, while the exponential characteristic of higher-order derivatives poses a risk of gradient explosion. Consequently, the 2nd-degree configuration is identified as the optimal engineering choice, providing a stable activation environment that avoids the exponential computational slowdown and mathematical volatility associated with higher-order approximations.

\subsection{Discussion}

The experimental results showed that the proposed approximation performed reliably across a range of experiments. In terms of MSE, the proposed Kernel  Polynomial method consistently achieved the lowest values in most scenarios, with the exception of situations with a higher variance in Experiment 1, where FasterCryptoNets performed better. However, the results of approximation accuracy for structured data revealed a significant limitation in the FasterCryptoNets approach, since it fails to account for negative values. In contrast, the Kernel Polynomial method provides a more robust and comprehensive approximation for general tokenized datasets since it successfully handles the full range of tokenized values, including negatives. This indicates that the Kernel-based Polynomial is both accurate and stable under typical input conditions.

When applied to DL models, the Kernel Polynomial approach provides predictive performance that is consistent across various architectures, such as FeedForward networks, CNNs, and ViT Transformers. Despite the fact that particular benchmarks, like FeedForward networks, indicate a slight advantage at the FasterCryptoNets, the suggested approach is still within the range of 0.3\% of the best result and has an great degree of resilience. In addition, the Kernel Polynomial emerges as the superior approximation as model complexity increases toward ViT architectures with the CIFAR-100 dataset. This performance indicates that whereas simpler approximations, such as X Squared, can be used to perform shallow tasks, the Kernel Polynomial method can be more easily scaled to high-dimensional feature space without the loss in accuracy, as in Chebyshev approximations. Notably, when moving from plaintext to the encrypted domain, the proposed approach maintains this consistency alongside low encrypted training time, at a low Accuracy Gap and Slowdown Factor, which can be invaluable in large-scale encrypted DL models.

In summary, through the evaluation process is demonstrated that the proposed Kernel Polynomial method offers a better option due to its lower MSE values, which confirmed the effectiveness of the approximation in capturing the target behavior with minimal deviation, its overall consistency and the unique advantages it provides, such as lower training time and higher model accuracy, as well as its continued effectiveness even when slightly outperformed. Considering its competitive performance alongside its lower approximation complexity and reduced computational overhead, the proposed Kernel Polynomial represents a robust and efficient activation approximation, considering the primary objective of this study in approximating ReLU with an HE-compatible polynomial.

\section{Conclusion}
\label{sec:Conclusion}

This work introduced a Kernel Polynomial approximation method of the ReLU activation function, designed to be efficient within HE frameworks that can address privacy concerns in NLP settings, while maintaining high accuracy. Unlike prior approaches, the proposed method was trained on real embedding vectors from pre-trained LLMs. Through systematic comparison with existing approximations identified through the literature, including X-Squared, FasterCryptoNets, and Chebyshev polynomials, the proposed method demonstrated significantly lower approximation error across various evaluation experiments, while achieving comparable predictive accuracy when integrated into both DL and transformer architectures, effectively bridging the performance gap between encrypted and unencrypted execution and secure inference in LLMs. 

The intriguing aspect of the assessment is that polynomials of higher degrees do not inherently guarantee a more precise approximation or better classification accuracy when integrated within NLP tasks. The practical demonstration showcases that for smooth functions, such as Kernel-based functions, exceptionally low approximation errors can be achieved even with lower-degree polynomials while maintaining high classification accuracy and lower computational costs.

Future research steps aim to overcome recognized limitations in high-variance contexts through the investigation of adaptive hybrid models. The implementation of the proposed Kernel Polynomial approximation method within LLMs will also be investigated, with an emphasis on enhancing processing speeds through hardware-level HE acceleration and coefficient optimization. Additionally, the resistance of the proposed Kernel Polynomial method to adversarial tokenized inputs will be further examined, since this is an essential requirement for implementation in high-security, privacy-preserving environments. Finally, to further illustrate the adaptability of the method in large-scale encrypted DL, its effectiveness will be evaluated across several data types, beyond image classification that was investigated in this study.

\bibliographystyle{ACM-Reference-Format}
\bibliography{references}

@Manual{kernlab,
  title  = {kernlab: Kernel-Based Machine Learning Lab},
  author = {Alexandros Karatzoglou and Alex Smola and Kurt Hornik},
  year   = {2024},
  note   = {R package version 0.9-33},
  url    = {https://CRAN.R-project.org/package=kernlab},
}

@misc{transformers,
  author = {Hugging Face},
  title = {Transformers: State-of-the-art Machine Learning for Pytorch, TensorFlow, and JAX},
  year = {2024},
  howpublished = {\url{https://huggingface.co/transformers}},
  note = {Accessed: 2025-05-23}
}

@misc{tenseal,
  author       = {OpenMined and Zama},
  title        = {TenSEAL: Homomorphic encryption library for PyTorch tensors},
  year         = 2024,
  howpublished = {\url{https://github.com/OpenMined/TenSEAL}},
  note         = {Accessed: 2025-05-23}
}

@inproceedings{10538562,
  author={Khan, Tanveer and Michalas, Antonis},
  booktitle={2023 IEEE 22nd International Conference on Trust, Security and Privacy in Computing and Communications (TrustCom)}, 
  title={Learning in the Dark: Privacy-Preserving Machine Learning using Function Approximation}, 
  year={2023},
  volume={},
  number={},
  pages={62-71},
  publisher = {IEEE Computer Society},
address = {Los Alamitos, CA, USA},
  doi={10.1109/TrustCom60117.2023.00031}}

@misc{juvekar2018faster,
      title={Faster CryptoNets: Leveraging Sparsity for Real-World Encrypted Inference}, 
      author={Edward Chou and Josh Beal and Daniel Levy and Serena Yeung and Albert Haque and Li Fei-Fei},
      year={2018},
      eprint={1811.09953},
      archivePrefix={arXiv},
      primaryClass={cs.CR},
      url={https://arxiv.org/abs/1811.09953},
}

@book{cheney1982approximation,
  title={Introduction to Approximation Theory},
  author={Cheney, E. W.},
  year={1982},
  publisher={AMS Chelsea Publishing},
  address ={201 Charles Street, Providence, RI 02904-2213, USA},
}

@book{pinkus1985nwidths,
  title={n-Widths in Approximation Theory},
  author={Pinkus, Allan},
  year={1985},
  publisher={Springer-Verlag},
  address = {Springer-Verlag Berlin Heidelberg 1985}
}

@inproceedings{socher2013recursive,
    title = "Recursive Deep Models for Semantic Compositionality Over a Sentiment Treebank",
    author = "Socher, Richard  and
      Perelygin, Alex  and
      Wu, Jean  and
      Chuang, Jason  and
      Manning, Christopher D.  and
      Ng, Andrew  and
      Potts, Christopher",
    editor = "Yarowsky, David  and
      Baldwin, Timothy  and
      Korhonen, Anna  and
      Livescu, Karen  and
      Bethard, Steven",
    booktitle = "Proceedings of the 2013 Conference on Empirical Methods in Natural Language Processing",
    month = oct,
    year = "2013",
    address = "Seattle, Washington, USA",
    publisher = "Association for Computational Linguistics",
    url = "https://aclanthology.org/D13-1170/",
    pages = "1631--1642"
}

@InProceedings{pmlr-v48-gilad-bachrach16,
  title = 	 {CryptoNets: Applying Neural Networks to Encrypted Data with High Throughput and Accuracy},
  author = 	 {Gilad-Bachrach, Ran and Dowlin, Nathan and Laine, Kim and Lauter, Kristin and Naehrig, Michael and Wernsing, John},
  booktitle = 	 {Proceedings of The 33rd International Conference on Machine Learning},
  pages = 	 {201--210},
  year = 	 {2016},
  editor = 	 {Balcan, Maria Florina and Weinberger, Kilian Q.},
  volume = 	 {48},
  series = 	 {Proceedings of Machine Learning Research},
  address = 	 {New York, New York, USA},
  month = 	 {20--22 Jun},
  publisher =    {PMLR},
  url = 	 {https://proceedings.mlr.press/v48/gilad-bachrach16.html}
}

@inproceedings{nair2010rectified,
  author =    {Vinod Nair and Geoffrey E. Hinton},
  title =     {Rectified Linear Units Improve Restricted {Boltzmann} Machines},
  booktitle = {Proceedings of the 27th International Conference on Machine Learning (ICML-10)},
  pages =     {807--814},
  year =      2010,
  editor =    {Johannes F{\"u}rnkranz and Thorsten Joachims},
  address =   {Haifa, Israel},
  month =     {June},
  URL =       {http://www.icml2010.org/papers/432.pdf},
  publisher = {Omnipress}
}

@book{scholkopf2002learning,
  title={Learning with Kernels: Support Vector Machines, Regularization, Optimization, and Beyond},
  author={Sch{\"o}lkopf, Bernhard and Smola, Alexander J.},
  year={2002},
  publisher={MIT Press},
  address={Cambridge, MA}
}

@misc{draper1998applied,
  title={Applied Regression Analysis},
  author={Draper, Norman R. and Smith, Harry},
  year={1998},
  publisher={John Wiley \& Sons}
}

@misc{cheon2017ckks,
      author = {Jung Hee Cheon and Andrey Kim and Miran Kim and Yongsoo Song},
      title = {Homomorphic Encryption for Arithmetic of Approximate Numbers},
      howpublished = {Cryptology {ePrint} Archive, Paper 2016/421},
      year = {2016},
      url = {https://eprint.iacr.org/2016/421}
}

@article{lecun1998gradient,
  title={Gradient-based learning applied to document recognition},
  author={LeCun, Yann and Bottou, Léon and Bengio, Yoshua and Haffner, Patrick},
  journal={Proceedings of the IEEE},
  volume={86},
  number={11},
  pages={2278--2324},
  year={1998},
  publisher={IEEE}
}

@article{mnist,
  title={The mnist database of handwritten digit images for machine learning research},
  author={Deng, Li},
  journal={IEEE Signal Processing Magazine},
  volume={29},
  number={6},
  pages={141--142},
  year={2012},
  publisher={IEEE}
}

@misc{wang2019glue,
      title={GLUE: A Multi-Task Benchmark and Analysis Platform for Natural Language Understanding}, 
      author={Alex Wang and Amanpreet Singh and Julian Michael and Felix Hill and Omer Levy and Samuel R. Bowman},
      year={2019},
      eprint={1804.07461},
      archivePrefix={arXiv},
      primaryClass={cs.CL},
      url={https://arxiv.org/abs/1804.07461}, 
}

@misc{devlin2019bert,
      title={BERT: Pre-training of Deep Bidirectional Transformers for Language Understanding}, 
      author={Jacob Devlin and Ming-Wei Chang and Kenton Lee and Kristina Toutanova},
      year={2019},
      eprint={1810.04805},
      archivePrefix={arXiv},
      primaryClass={cs.CL},
      url={https://arxiv.org/abs/1810.04805}, 
}

@misc{ahle2020oblivioussketchinghighdegreepolynomial,
      title={Oblivious Sketching of High-Degree Polynomial Kernels}, 
      author={Thomas D. Ahle and Michael Kapralov and Jakob B. T. Knudsen and Rasmus Pagh and Ameya Velingker and David Woodruff and Amir Zandieh},
      year={2020},
      eprint={1909.01410},
      archivePrefix={arXiv},
      primaryClass={cs.DS},
      url={https://arxiv.org/abs/1909.01410}, 
}

@misc{Krizhevsky2009LearningML,
  title={Learning Multiple Layers of Features from Tiny Images},
  author={Alex Krizhevsky},
  year={2009},
  url={https://api.semanticscholar.org/CorpusID:18268744}
}

@article{FORNBERG2007379,
title = {The Runge phenomenon and spatially variable shape parameters in RBF interpolation},
journal = {Computers \& Mathematics with Applications},
volume = {54},
number = {3},
pages = {379-398},
year = {2007},
issn = {0898-1221},
doi = {https://doi.org/10.1016/j.camwa.2007.01.028},
url = {https://www.sciencedirect.com/science/article/pii/S0898122107002210},
author = {Bengt Fornberg and Julia Zuev},
}

\end{document}